\documentclass[onecolumn,12pt]{IEEETran}
\usepackage{bbm}
\usepackage{amsfonts}
\usepackage{tipa}
\usepackage{amssymb}
\usepackage{mathrsfs}
\usepackage{graphicx}
\usepackage{color, soul}
\usepackage{amsfonts,amssymb,amsmath}
\usepackage{latexsym}
\usepackage{multirow}
\usepackage{hyperref}
\usepackage{array}
\usepackage{epsfig,epstopdf}
\usepackage{array}

\newtheorem{lemma}{Lemma}

\begin{document}
\title{Fading Two-Way Relay Channels: Physical-Layer Versus Digital Network Coding}
\author{Zhi~Chen~\IEEEmembership{Member,~IEEE,} Teng~Joon~Lim~\IEEEmembership{Senior Member,~IEEE,}
        and~Mehul~Motani~\IEEEmembership{Member,~IEEE}
\thanks{Part of this work was accepted by IEEE Globecom 2013. This work was partially funded by grant R-263-000-649-133 and R-263-000-579-112 from the Ministry of Education
Academic Research Fund.

The authors are with the Department of Electrical and Computer Engineering, National University of Singapore, Singapore, 117583. Tel: +65 6601 2055.
Fax: +65 6779 1103.
Emails: elecz@nus.edu.sg; eleltj@nus.edu.sg; motani@nus.edu.sg.}}


\maketitle

\maketitle
\baselineskip 24pt
\begin{abstract}\\
\baselineskip=18pt
In this paper, we consider three transmit strategies for the fading three-node,
two-way relay network (TWRN) -- physical-layer network coding (PNC),
digital network coding (DNC) and codeword superposition (CW-Sup).
The aim is to minimize the total average energy needed to deliver a given pair of required average rates.
Full channel state information is assumed to be available at all transmitters and receivers.
The optimization problems corresponding to the various strategies in fading channels are formulated, solved and compared.
For the DNC-based strategies, a simple time sharing of transmission of the network-coded message and the remaining bits of the larger message (DNC-TS) is considered first. We extend this approach to include a superposition strategy (DNC-Sup), in which the network-coded message and the remainder of the longer source message are superimposed before transmission. It is demonstrated theoretically that DNC-Sup outperforms DNC-TS and CW-Sup in terms of total average energy usage.
More importantly, it is shown in simulation that DNC-Sup performs better than PNC
if the required rate is low and worse otherwise.
Finally, an algorithm to select the optimal strategy
in terms of energy usage subject to different rate pair requirements
is presented.

\baselineskip=18pt
\end{abstract}

\IEEEpeerreviewmaketitle

\section{Introduction}
Network coding, introduced in \cite{ahlswede2000network}, has proven to be an important tool for improving network throughput.
As one of the simplest applications of network coding, the two-way relay network (TWRN) has been extensively studied in the literature. A TWRN usually
comprises two source nodes ($S_1$ and $S_2$) and one relay node
($R$), where $S_1$ and $S_2$ have information to exchange with each
other. A direct link between $S_1$ and $S_2$ is unavailable.

In the literature, several transmission methods have been proposed for the TWRN,
such as digital network coding (DNC) in \cite{liu2008network}--\cite{Chen2012two-way6},
codeword superposition (CW-Sup) in \cite{Rankov} and \cite{Boche}, physical-layer network coding (PNC)
in \cite{Nam}--\cite{popovski2007physical} and analog network coding
(ANC) in \cite{avestimehr2010capacity}--\cite{Katti}.
Among them, DNC requires the relay node to jointly decode individual messages from both $S_1$ and $S_2$ on
the uplink and then combine them together as a new message for delivery on the downlink.
The codeword superposition strategy superimposes the two individual codewords received on the
uplink and forwards the result to both sources. PNC requires the relay node
to decode a function of the two messages instead of the two individual messages and
forward this function message to both sources.
Another method, ANC, requires the relay node to
simply amplify the mixed signals received over the uplink without decoding, and
 forward this amplified signal to the source nodes.

\begin{figure}[!h]
   \centering
   \includegraphics[width = 12cm]{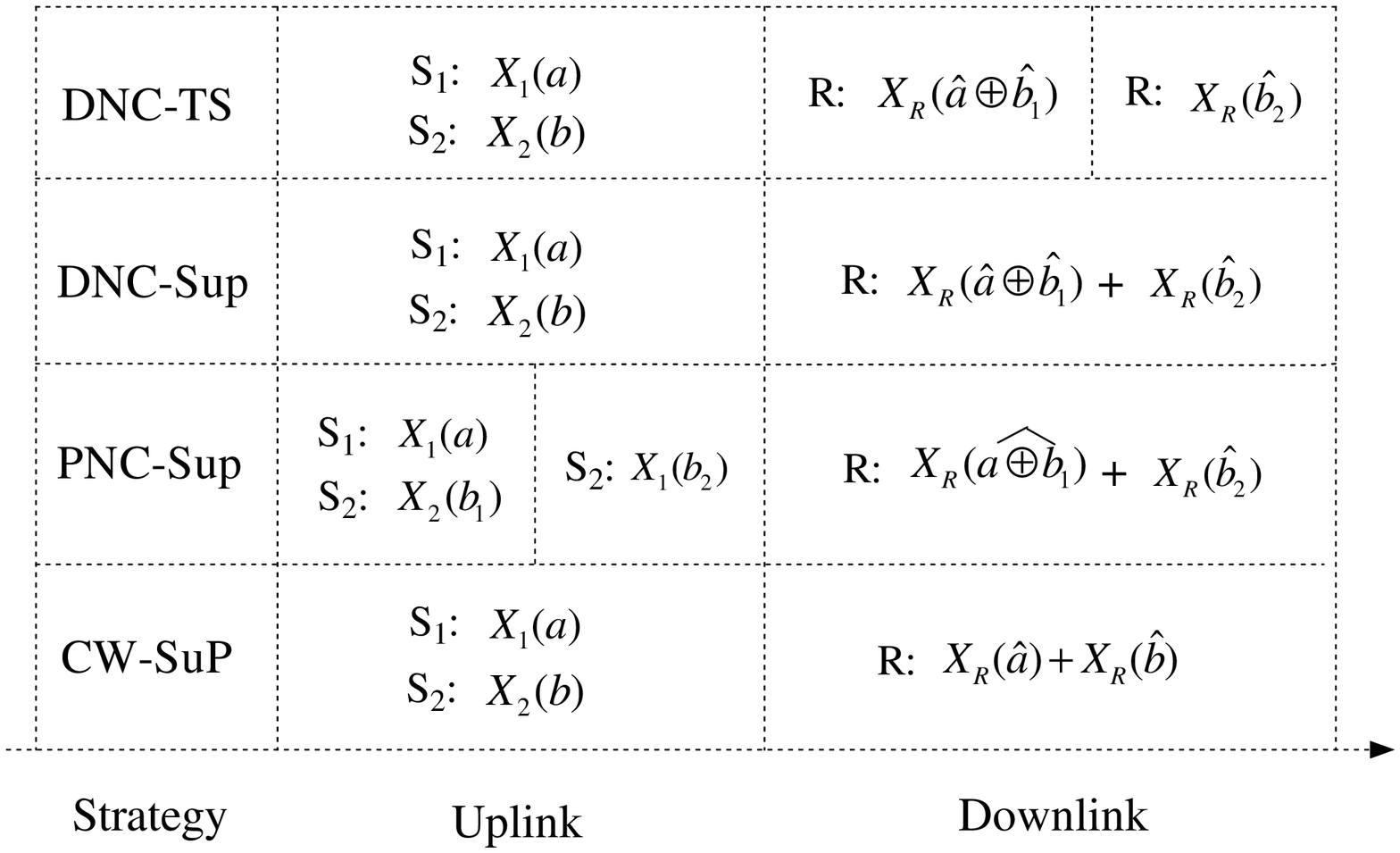}
   \caption{Illustration of different strategies considered.} \label{fig:Comparison_NC}
   \end{figure}

In all cases, since each source node has perfect
 knowledge of the message originating from itself, it can subtract its own transmitted
 message and obtain the intended message from the other source upon receipt of the
 relay's network-coded message. Note that ANC performs worse than PNC as noise at the
 relay is amplified when transmitting on the downlink, thus degrading the achievable rate, as was shown in \cite{nazer2011physical}.
 Hence in this work we shall only consider PNC, DNC and CW-Sup strategies over fading channels.
 For clarity, a graphical description of the strategy considered is depicted in
 Fig.\ \ref{fig:Comparison_NC}, {where the message transmitted by $S_1$
 is denoted by $a$,
 and the one from $S_2$ is $b$. Without loss of generality,
 we assume that the message $b$ is
 longer than $a$.
 It is composed of two parts: $b_1$ and $b_2$,
 where $b_1$ is assumed to have the same length as the message $a$.
 With the DNC based strategy or CW-Sup,
 the decoded messages at $R$ are $\hat{a}$ and $\hat{b}$.
 With PNC-Sup, the relay decodes
 the function message $\widehat{a \oplus b_1}$ and the remaining bits $\hat{b_2}$.}
 It should be noted that PNC is a family of techniques, and the PNC strategy considered in this work is a specific one from
 \cite{wilson2010joint}. In addition, in some works in the literature,
 network coding
 schemes with a conventional multi-access uplink for TWRNs were also referred to
 as PNC schemes. In this work, however, to distinguish a conventional
 multi-access uplink from
 the idea of decoding a function message over the uplink,
 we shall refer to the former as a DNC scheme.

In a TWRN, an achievable rate region with DNC was first investigated in \cite{liu2008network} for a three-slot protocol. In \cite{fong2011practical}, an achievable rate region with only time-resource allocation was investigated for the case of static channels.
In \cite{Boche}, a codeword superposition technique was discussed and only the optimal
time division between the uplink and the downlink phases was investigated.
In \cite{Nam} and \cite{wilson2010joint},
the capacity region employing PNC was derived for AWGN channels without any discussion of resource
allocation. In \cite{Tarokh} the optimized
constellation for TWRNs with PNC was investigated for
symmetric traffic scenario, by adjusting the PNC map with the channel gains.
In \cite{Katti}, analog network coding is studied as a way to utilize interference in wireless networks.
Furthermore, PNC was demonstrated to be better than DNC with time sharing uplink in \cite{Liew2006} in terms of achievable throughput as long as SNR is higher than -5dB.

In \cite{Chen2012two-way}, \cite{Chen2012two-way5} and \cite{Chen2012two-way6}, we investigated the minimization of total energy usage for a three-node TWRN subject to stability constraints,
with various allowed transmit modes based on DNC.
Only an orthogonal, time-sharing uplink, as well as time sharing of the digital network coded bits with
the remaining bits in the downlink were considered in \cite{Chen2012two-way}. In \cite{Chen2012two-way5},
joint decoding on the uplink was allowed, ensuring that all points in the multi-access channel
rate region are achievable. The non-fading (static) case was considered in \cite{Chen2012two-way5}.
The fading scenario was investigated in the preliminary version of this work in \cite{Chen2012two-way6}.
However, no mention of PNC and CW-Sup was made in \cite{Chen2012two-way}--\cite{Chen2012two-way6}.
It is also noted that the strategies considered in this work are different from
the traditional three-phase DNC and two-phase PNC. The DNC schemes
considered in this paper utilize a multi-access uplink, while the three-phase DNC
adopts the time sharing uplink transmission. In addition, the traditional DNC or
PNC usually considers the symmetric traffic scenario, whereas in this work we assume asymmetric traffic.

Although in \cite{Liew2006} PNC was shown to perform better than DNC with a time-sharing uplink,
its performance relative to
DNC with a multi-access uplink is still unknown.
In this work, we will focus on comparing the performance and complexity of the optimized PNC,
DNC and CW-Sup strategies, with a multi-access uplink assumed for the latter two.
It will be shown that the proposed PNC strategy outperforms other strategies with relatively high data rate requirements.
In the regime of low data rate requirements, DNC-Sup performs better than
the proposed PNC and CW-Sup strategies.
To summarize, our main contributions are:
\begin{itemize}
\item We find the resource allocation that minimizes the total average energy required to support a given
rate requirement in a three-node TWRN in a fading channel.
\item We prove that DNC-Sup outperforms DNC-TS and CW-Sup under all channel conditions.
\item We show that the proposed PNC-Sup strategy outperforms DNC-Sup in applications with high data rate requirements on both sides.
However, if data rate requirements on both sides are low, DNC-Sup is preferred in terms of total average transmit energy usage.
\item Based on the analysis of different schemes,
we present an optimal algorithm to select the
best strategy with different rate pair requirements in terms of energy usage.
\end{itemize}

The rest of this paper is organized as follows. In Section II, we describe the three-node TWRN and our setup.
In Section III-A, we briefly describe the strategies considered.
From Section II-B to Section II-E, we discuss respectively
the PNC strategy, the orthogonal time-sharing of DNC message and
the excess bits of the longer message,
the superposition of the DNC bits and the excess bits of the longer
message, and the superposition of the source messages.
In Section III-F, an optimal algorithm to always
select the best strategy is presented.
Numerical results are presented in Section IV and
Section V concludes this work.


\section{System Description}
We consider a three-node, two-way relay network consisting of two sources $S_1$, $S_2$ and one relay node $R$.
$S_1$ and $S_2$ exchange information through the relay $R$ without a direct link between them.
A block flat-fading channel model is assumed for all links, i.e., the channel is a constant over one slot, defined as a cycle through all defined transmission modes in a strategy.
The instantaneous channel power gains, corresponding to links $S_1$-$R$, $S_2$-$R$, $R$-$S_1$ and $R$-$S_2$, are defined as $g_{1r}$, $g_{2r}$, $g_{r1}$ and $g_{r2}$, respectively,
their averages by $\bar{g}_{ij}$ ($i,j=1,2,r$),
and their probability density
functions by $p(g_{ij})$. We also assume ergodicity in the channel processes. Noise at every node is modeled by an i.i.d.\ Gaussian random variable with zero
mean and unit variance. Each node is equipped with one antenna and works in
half-duplex mode, i.e., it cannot transmit and receive simultaneously.
It is assumed that all transmitters incur some energy overhead
due to the energy needed to operate the transmitter hardware, which is a constant independent
of transmit power. This energy overhead is defined as $P_Z$ for all three nodes together.

For each transmission strategy, the following optimization is performed. Given instantaneous channel state information (CSI) as well as their probability distributions, find
\begin{enumerate}
  \item the optimal time fraction to allocate to each mode in the transmission strategy over all time, i.e., the designed values are applied regardless of instantaneous CSI, and
  \item the channel-dependent optimal power and rate allocations for each node, in each mode.
\end{enumerate}
Here, the optimal solution is the one that minimizes the total energy used. The constraints that must be met are
that the {\em average} rate from $S_1$ to $S_2$ is at least $\lambda_1$, and that from $S_2$ to $S_1$
is at least $\lambda_2$.

\section{Transmission Strategies}

\subsection{Brief Description}
In this work, we consider four strategies: PNC-Sup, DNC-TS, DNC-Sup and CW-Sup, as shown in Fig. \ref{fig:Comparison_NC}. Without loss of generality, we let $\lambda_1 \leq \lambda_2$ in the following analysis.

In the PNC-Sup strategy, there are a total of three transmission modes.
The first mode is when $S_1$ and $S_2$ simultaneously transmit
two messages ($a$ and $b_1$ respectively) of the same length
at the same rate, encoded so that $R$ is able to decode the sum $a \oplus b_1$ (see \cite{wilson2010joint} for details).
In the second mode, with $\lambda_1 \leq \lambda_2$,
$S_2$ transmits to $R$ the bits
that were not transmitted in the first mode ($b_2$),
at a rate that will be found through solving the
optimization problem below.
Finally, in the third mode, the relay node superimposes the sum
message ($a \oplus b_1$) on $b_2$,
and broadcasts the combination to $S_1$ and $S_2$.

In DNC-TS, we consider a multi-access uplink
and time sharing of the digital network-coded message (denoted as $a\oplus b_1$ in Fig. \ref{fig:Comparison_NC})
and the remaining bits of the longer message ($b_2$) on the downlink.
On the other hand, in DNC-Sup, a multi-access uplink and the
superposition of the digital network coded bits and the
remaining bits of the longer message on the downlink are considered.

The last strategy considered is taken from \cite{Rankov}, CW-Sup,
and comprises a multi-access uplink and a codeword
superposition of the two original messages from $S_1$ and
$S_2$ at $R$ for transmission on the downlink. No explicit network coding technique is applied in this strategy.

Note that from queueing theory, for a queueing system with
the service rate close to the arrival rate,
there are packets buffered at all nodes
with high probability.
To make the problem tractable for PNC-Sup (inherently required by the second mode of PNC-Sup), we further introduce the constraint that
if $\lambda_1 \le \lambda_2$, then the message to be transmitted in each slot from $S_1$ is shorter
than the one from $S_2$, irrespective of the instantaneous channel gains, and vice versa for the
case $\lambda_1 > \lambda_2$.

Note also that for all strategies if $\lambda_1 \le \lambda_2$, with
the service rate close to the arrival rate, the number of bits from $S_1$ buffered at $R$ is
smaller than that from $S_2$ with high probability.
Hence we assume that it can always transmit
more data to $S_1$
in each slot, irrespective of the instantaneous channel gains, and vise versa for the case $\lambda_1 > \lambda_2$.

\subsection{Physical Layer Network Coding Strategy (PNC-Sup)}

In this strategy, there are three transmission modes.
In mode one, $S_1$ and $S_2$ transmit to $R$ $a$ and $b_1$ of the same length, respectively, and R decodes $a+b_1$. In mode two,
$S_2$ transmits $b_2$ to $R$. In mode three,
the relay broadcasts to $S_1$ and $S_2$ the superposition of $\widehat{a+b_1}$
and $\hat{b_2}$.

For the first mode, it was demonstrated
in \cite{wilson2010joint} that the achievable symmetric transmit rate with
PNC is $\log_2(1/2+SNR)$ over the AWGN channel\footnote{{Note that
research on the achievable rate of PNC is still an open topic.
Although an ideal exchange rate of $1/2 \log_2 (1 + SNR)$ is suggested in [15],
the achievable PNC rate
exploits the recent result given in [17].}},
where $SNR$ is the signal-to-noise ratio at the receiver from both $S_1$ and $S_2$.
The details of how to practically transmit rates close to this theoretical capacity are given
in \cite{wilson2010joint}.
{The $\log_2(1/2 + SNR)$ expression implies that transmit powers at
$S_1$ and $S_2$ are adjusted at each channel use so that the instantaneous
SNR in the $S_1$-$R$ and $S_2$-$R$ links are identical.
While this is sub-optimal, no other simple capacity expression exists
for PNC and therefore we use this scheme in this paper.
Note that the PNC map is matched to channel fading coefficients in [10],
but no rate expression was derived.}
The power required to transmit a message at rate $R_{1i}^{\mathrm{PNC}}$ from $S_i$ in the first mode
is therefore given by,
\begin{align}
P_{1i}^{\mathrm{PNC}}(g_{ir})=\frac{2^{R_{1i}^{\mathrm{PNC}}}-\frac{1}{2}}{g_{ir}}. \label{eqn:pnc1}
\end{align}
The total power required in the first mode then is,
\begin{align}
P_{1}^{\mathrm{PNC}}(g_{1r},g_{2r})=\sum_{i=1}^2 \frac{2^{R_{1i}^{\mathrm{PNC}}}-\frac{1}{2}}{g_{ir}}.\label{eqn:pnc2}
\end{align}
The associated transmit rate $R_1^{\mathrm{PNC}}=R_{1i}^{\mathrm{PNC}}$ is given by,
 \begin{align}
 R_1^{\mathrm{PNC}}=\log_2 \left(\frac{1}{2}+P_{11}^{\mathrm{PNC}}g_{1r} \right)=\log_2 \left(\frac{1}{2}+P_{12}^{\mathrm{PNC}}g_{2r} \right),
 \end{align}
 where the channel inversion equality $P_{11}^{\mathrm{PNC}}g_{1r}=P_{12}^{\mathrm{PNC}}g_{2r}$
 is required [17] for the validity of the PNC rate expressions.

 In the second mode, the remaining bits of $S_2$ will be delivered to the relay node.
 The minimum power needed by $S_2$ to transmit its remaining bits at the rate $R_{22}^{\mathrm{PNC}}$ is
 \begin{align}
 P_{22}^{\mathrm{PNC}}(g_{2r})=\frac{2^{R_{22}^{\mathrm{PNC}}}-1}{g_{2r}},
 \end{align}
which comes from the Shannon channel capacity for point-to-point Gaussian
channels, i.e., $R_{22}^{\mathrm{PNC}}=\log_2\left(1+P_2^{\mathrm{PNC}}(g_{2r}) g_{2r})\right)$.
The total transmit power then is
$P_{2}^{\mathrm{PNC}}=P_{22}^{\mathrm{PNC}}$ and the total transmit rate
$R_{2}^{\mathrm{PNC}}=R_{22}^{\mathrm{PNC}}$.

The power required in the downlink phase (the third mode) consists of two parts.
One is to transmit the common network-coded message at rate $R_{3,c}^{\mathrm{PNC}}(g_{r1},g_{r2})$
and the other is to transmit the remaining bits of the message from $S_2$ to $S_1$ at
rate $R_{3,p}^{\mathrm{PNC}}(g_{r1},g_{r2})$. The decoding method at $S_1$ and $S_2$ is as follows. Firstly, both source nodes decode the common
message in the presence of the interference from the private message. Hence $S_1$ can subtract its own message from the
network-coded message and obtain part of the message from $S_2$. $S_2$ can
also subtract its own message from the network-coded message and obtain the
full message from $S_1$. $S_1$, however, needs to decode
the remaining bits of the larger message from $S_2$, and then combine it with the message
embedded in the network-coded message to obtain the full message from $S_2$.
Hence, from the Shannon capacity formula,
\begin{align}
R_{3,c}^{\mathrm{PNC}}&=\min_i \log_2\left(1+\frac{P_{3,c}^{\mathrm{PNC}}(g_{r1},g_{r2}) g_{ri}}{1+P_{3,p}^{\mathrm{PNC}}(g_{r1},g_{r2})g_{ri}}\right) \label{eqn:bc1}\\
&=\log_2\left(1+\frac{P_{3,c}^{\mathrm{PNC}}(g_{r1},g_{r2}) \min( g_{r1},g_{r2})}{1+P_{3,p}^{\mathrm{PNC}}(g_{r1},g_{r2})\min(g_{r1},g_{r2})}\right),\label{eqn:bc2}
\end{align}
where $P_{3,c}^{\mathrm{PNC}}(g_{r1},g_{r2})$ is the power required for transmission of the network-coded bits
and $P_{3,p}^{\mathrm{PNC}}(g_{r1},g_{r2})$ is that for the remaining bits of the larger message.
(\ref{eqn:bc2}) comes from the fact that $P_{3,c}^{\mathrm{PNC}} g/(1+P_{3,p}^{\mathrm{PNC}}g)$ is a monotonically
 increasing function of $g$ if $g>0$.
The power required to transmit the network-coded message then is given by,
\begin{align}
P_{3,c}^{\mathrm{PNC}}(g_{r1},g_{r2})= \frac{\left(2^{R_{3,c}^{\mathrm{PNC}}}-1\right)\left( 1+P_{3,p}^{\mathrm{PNC}}\min(g_{r1},g_{r2})  \right) }{\min (g_{r1},g_{r2})}.
\end{align}

For the remaining bits after subtracting the network-coded message, the achievable rate is given by
$$R_{3,p}^{\mathrm{PNC}}=\log_2\left(1+P_{3,p}^{\mathrm{PNC}}(g_{r1},g_{r2})g_{r1}\right),$$
and the power required by the relay to transmit at this rate is
\begin{align}
P_{3,p}^{\mathrm{PNC}}(g_{r1})= \frac{\left(2^{R_{3,p}^{\mathrm{PNC}}}-1\right)}{g_{r1}}.
\end{align}
The total power required in the third mode is given by,
\begin{align}
{P}_3^{\mathrm{PNC}}=P_{3,c}^{\mathrm{PNC}}+P_{3,p}^{\mathrm{PNC}}.
\end{align}

We define $\bar{P}_i^{\mathrm{PNC}}$ and $\bar{R}_i^{\mathrm{PNC}}$ as the average transmit power required
and the associated average transmit rate for the $i$th phase, respectively,
where the expectation is over the associated channel distributions.
In addition, $f_i^{\mathrm{PNC}}$ ($i=1,2,3$) denotes the time fraction assigned to the $i$th phase.
$f_i^{\mathrm{PNC}}\bar{P}_i^{\mathrm{PNC}}$ is proportional to the average transmit energy consumed in the $i$th phase.

Hence, assuming that $\lambda_1 \le \lambda_2$,
minimizing average energy-usage with average rate constraints and PNC at the relay, is formulated as follows. We call this problem {\bf P1}.
\begin{align}
\min_{f_i,R_i^{\mathrm{PNC}}(g_{ij})} \quad \sum_{i=1}^3 f_i^{\mathrm{PNC}}\bar{P}_i^{\mathrm{PNC}}+P_Z \label{eqn:pncobj}
\end{align}
subject to:
\begin{align}
f_1^{\mathrm{PNC}}\bar{R}_1^{\mathrm{PNC}} &\ge \lambda_1 \label{eqn:pnccon1}\\
f_2^{\mathrm{PNC}}\bar{R}_{22}^{\mathrm{PNC}} &\ge \lambda_2-\lambda_1 \label{eqn:pnccon2}\\
f_3^{\mathrm{PNC}}\bar{R}_{3,c}^{\mathrm{PNC}} &\ge \lambda_1 \label{eqn:pnccon3}\\
f_3^{\mathrm{PNC}}\bar{R}_{3,p}^{\mathrm{PNC}} &\ge \lambda_2-\lambda_1 \label{eqn:pnccon4}\\
P_{11}^{\mathrm{PNC}}g_{1r}&=P_{12}^{\mathrm{PNC}}g_{2r} \label{eqn:pnccon5}\\
\sum_{i=1}^3 f_i^{\mathrm{PNC}}  &\leq 1   \label{eqn:pnccon6}
\end{align}
where the target function in (\ref{eqn:pncobj}) is the average energy consumed per slot. $\lambda_1$ (assuming $\lambda_1 \leq \lambda_2$) in (\ref{eqn:pnccon1})
comes from the fact that only part of the larger message is simultaneously transmitted with the
smaller message such that the two messages have the same length. $\lambda_2-\lambda_1$ in (\ref{eqn:pnccon2}) is for unicast transmission of the remaining bits of the larger message.
(\ref{eqn:pnccon5}) comes from the channel-inversion equality
to alleviate intrinsic interference.
Note that the overhead energy $P_Z$ is a constant and its value does not affect the optimal solution to {\bf P1}.
Hence we simply assume $P_Z=0$ in the rest of this work.

Note that the objective function
is a convex function of the transmit rates and a linear function of the
time fraction of each mode.
In addition, the constraints are also linear functions of time
fractions and/or of the transmit rates.
Therefore, {\bf P1} is a standard convex optimization problem and
it can be solved by
the Lagrange multiplier method \cite{boyd2004convex}.
By taking the first-order derivative of the Lagrangian with respect to the related parameters, the associated KKT conditions are given by,
\begin{align}
\bar{P}^{\mathrm{PNC}}_i-\beta_i\bar{R}^{\mathrm{PNC}}_i&=\gamma, \quad i=1,2 \label{eqn:pnckkt1} \\
\bar{P}^{\mathrm{PNC}}_3-\beta_{3,c}\bar{R}^{\mathrm{PNC}}_{3,c}-\beta_{3,p}\bar{R}^{\mathrm{PNC}}_{3,p} &=\gamma, \label{eqn:pnckkt2}\\
2^{R_1^{\mathrm{PNC}}}\ln2 \left(\frac{1}{g_{1r}}+\frac{1}{g_{2r}}\right)-\beta_1 &= 0, \label{eqn:pnckkt3}\\
\frac{2^{R_2^{\mathrm{PNC}}}}{g_{2r}} \ln2 -\beta_2 &= 0, \label{eqn:pnckkt4}
\end{align}
and constraints in (\ref{eqn:pnccon1})-(\ref{eqn:pnccon4}) and (\ref{eqn:pnccon6}) are satisfied with equality.
The Lagrangian multiplier $\beta_i$ is with the rate requirement
in the $i$th phase and $\gamma$ with
the physical constraint in (\ref{eqn:pnccon6}).


From these KKT conditions, the associated optimal power allocation strategy is given by
\begin{align}
P_{1}^{\mathrm{PNC}}(g_{1r},g_{2r})&=\left[\beta_1^* \log_2 e - \frac{1}{2} \left( \frac{1}{g_{1r}}+\frac{1}{g_{2r}}  \right)\right]^+ , \label{eqn:pncopt1}\\
P_{1i}^{\mathrm{PNC}}(g_{1r},g_{2r})&=\left[\beta_1^* \log_2 e \frac{g_{3-i,r}}{g_{1r}+g_{2r}}-\frac{1}{2g_{ir}}\right]^+, \label{eqn:pncopt2}\\
P_2^{\mathrm{PNC}}(g_{r1},g_{r2})&=\left[\beta_2^*\log_2 e - \frac{1}{g_{2r}}\right] ,\label{eqn:pncopt3}
\end{align}
where the asterisks denote optimality. $[\cdot]^+=\max(\cdot, 0)$ and (\ref{eqn:pncopt2})
comes from (\ref{eqn:pnc1}), (\ref{eqn:pnc2}) and (\ref{eqn:pncopt1}). (\ref{eqn:pncopt3})
comes from the assumption that $\lambda_1 \leq \lambda_2$. It is observed that the total
power allocation in the uplink, i.e., $P_{1}^{\mathrm{PNC}}(g_{1r},g_{2r})$, has the
water-filling structure in (\ref{eqn:pncopt1}), with the instantaneous channel gains $g_{1r}$
as well as $g_{2r}$ taken into account.

By deriving the first-order derivative of (\ref{eqn:pnckkt2}) (similar to power allocation for broadcast channels in \cite{David}), the optimal power allocation in the downlink is summarized in Lemma \ref{downlink1}.
\begin{lemma} \label{downlink1}
If $g_{r1}>g_{r2}$, the power allocation
 in the downlink is
\begin{itemize}
\item{} if $\beta_{3,p}^*g_{r1} \leq \beta_{3,c}^*g_{r2}$, then
\begin{align}
\left\{
\begin{array}{ll}
P_{3,c}^{\mathrm{PNC}}=[\beta_{3,c}^*\log_2e-\frac{1}{g_{r2}}]^+\\
P_{3,p}^{\mathrm{PNC}}=0\\
\end{array}
\right.
\end{align}
\item{} if $\beta_{3,p}^*g_{r1}>\beta_{3,c}^*g_{r2}$ and $(\beta_{3,c}^*-\beta_{3,p}^*)\log_2e \leq \frac{g_{r1}-g_{r2}}{g_{r1}g_{r2}}$, then
\begin{align}
\left\{
\begin{array}{ll}
P_{3,c}^{\mathrm{PNC}}=0\\
P_{3,p}^{\mathrm{PNC}}=[\beta_{3,p}^*\log_2e-\frac{1}{g_{r1}}]^+\\
\end{array}
\right.
\end{align}
\item{} if $\beta_{3,p}^*g_{r1}>\beta_{3,c}^*g_{r2}$ and $(\beta_{3,c}^*-\beta_{3,p})\log_2e > \frac{g_{r1}-g_{r2}}{g_{r1}g_{r2}}$, then
\begin{align}
\left\{
\begin{array}{ll}
P_{3,c}^{\mathrm{PNC}}=[\beta_{3,c}^*\log_2 e-\frac{(\beta_{3,p}^*g_{r1}-\beta_{3,c}^*g_{r2})}{(\beta_{3,c}^*-\beta_{3,p}^*)g_{r1}g_{r2}}]^+\\
P_{3,p}^{\mathrm{PNC}}=[\frac{(\beta_{3,p}^*g_{r1}-\beta_{3,c}^*g_{r2})}{(\beta_{3,c}^*-\beta_{3,p}^*)g_{r1}g_{r2}}]^+\\
\end{array}
\right.
\end{align}
\end{itemize}
For the case that $g_{r1}<g_{r2}$ power allocation for the third mode in the
downlink can be derived in a similar manner and is omitted for brevity.
\end{lemma}

A multi-bisection method is used to numerically find the optimal solution to {\bf P1}. The procedure is as follows.
\begin{enumerate}
\item For a given $f_1^{\mathrm{PNC}}$ we can obtain $\bar{R}_1^{\mathrm{PNC}}$ from (\ref{eqn:pnccon1}) and then find the appropriate $\beta_{1}$ and $\bar{P}_1^{\mathrm{PNC}}$ from (\ref{eqn:pncopt1}) and (\ref{eqn:pncopt2}). We then compute $\gamma$ from (\ref{eqn:pnckkt1}).

\item With the computed $\gamma$, we can then iteratively find the appropriate $\beta_2$, $\beta_{3,c}$ and $\beta_{3,p}$ for the second mode and the third mode
satisfying (\ref{eqn:pnckkt1}) and (\ref{eqn:pnckkt2}) subject to the pre-defined accuracy, and obtain the
associated $\bar{R}_i^{\mathrm{PNC}}$, $\bar{P}_i^{\mathrm{PNC}}$ and $f_i^{\mathrm{PNC}}$ from (\ref{eqn:pnccon3})-(\ref{eqn:pnccon5}), (\ref{eqn:pncopt3}) and Lemma \ref{downlink1}. For instance, the pre-defined accuracy for Mode 2 can be $|\bar{P}^{\mathrm{PNC}}_2-\beta_i\bar{R}^{\mathrm{PNC}}_2-\gamma|<\epsilon_1$ where $\epsilon_1$ is a very small number.

\item We hence iterate $f_1^{\mathrm{PNC}}$ with all the computed $f_i^{\mathrm{PNC}}$
and follow the same procedure in 1) and 2) until we arrive at the optimal solution subject to the pre-defined accuracy,
e.g., the convergence constraint can be $1-\epsilon_2<\sum_{i=1}^3 f_i^{\mathrm{PNC}}<1$, where $\epsilon_2$ is also a very small positive number.
\end{enumerate}
Note that $\epsilon_1<\epsilon_2$ must be satisfied since the inner iteration should have higher accuracy
than the outer iteration in numerical computation, e.g., we can set $\epsilon_1=10^{-6}$ and $\epsilon_2=10^{-3}$.
This algorithm is presented to obtain the optimal Lagrangian multipliers. In simulation,
it is shown to converge to the global optimal solution in only tens of iterations.

\subsection{Digital Network Coding and Time Sharing in the Downlink (DNC-TS)}\label{time sharing}
In this section, we consider the use of digital network coding in
place of PNC,
and seek similarly to minimize total average energy usage.
The downlink time-shares a network coded message and a message of the remaining bits from one source,
and explains why we call this strategy DNC-TS.
The four modes in this strategy are described as follows.
\begin{itemize}
\item {\em Mode 1:} $S_1$ and $S_2$ simultaneously transmit to $R$ at average rates $\bar{R}_{11}$ and rate $\bar{R}_{12}$ with average powers $\bar{P}_{11}$ and $\bar{P}_{12}$ in the multi-access uplink respectively.
\item {\em Mode 2:} $R$ broadcasts to $S_1$ and $S_2$ at an average rate $\bar{R}_{2}$
with an average power $\bar{P}_{2}$ using digital network coding.
\item {\em Mode 3:} $R$ transmits only to $S_1$ at an average rate $\bar{R}_{3}$ with
an average power $\bar{P}_{3}$.
\item {\em Mode 4:} $R$ transmits only to $S_2$ at an average rate $\bar{R}_{4}$ with
an average power $\bar{P}_4$.
\end{itemize}


In order for the subscripts
to match the transmission modes, we introduce the new definitions $g_{11} = g_{1r}$, $g_{12} = g_{2r}$,
$g_2 = \min(g_{r1},g_{r2})$, $g_3 = g_{r1}$ and $g_4 = g_{r2}$ for each mode.


In DNC-TS, Mode 3 (if $S_2$ transmits the longer message)
and Mode 4 (if $S_1$ transmits the longer message) are useful
for transmitting bits that cannot be network-coded due to the
asymmetric message sizes. Without them, we would have to zero-pad
the shorter message in order to apply network coding in Mode 2.
However, since the network-coded message must be decoded by {\em both}
sources, the rate in Mode 2 is constrained by the smaller of $g_{r1}$
and $g_{r2}$. In contrast, the message in Mode 3 is only for $S_1$,
and therefore its rate is limited only by $g_{r1}$, and similarly
for Mode 4. It is thus always beneficial, in the asymmetric
case of interest here, to use Mode 3 or 4, in addition to the
network-coded Mode 2.

Let $P_{1i}(g_{11}, g_{12})$ ($i = 1, 2$) be the transmit power of $S_i$ for a given gain pair ($g_{11}$,$g_{12}$). For the multi-access uplink transmit mode, i.e., Mode 1, we have
$\bar{P}_1=\sum_{i=1}^2\bar{P}_{1i}$ and
\begin{align}
\bar{P}_{1i}= E \{P_{1i}(g_{11},g_{12})\},
\end{align}
where the expectation is taken over the distribution of channel gains.
As stated in \cite{David}, the optimal decoding order for a multi-access channel is to firstly decode the data from the stronger user, i.e.,
the user with the better uplink channel gain. This result follows from
MAC-BC duality. Hence with the assumption that $g_{11}<g_{12}$, we have
\begin{align}
R_{11}& =\log_2(1+P_{11}g_{11}),  \label{eqn:MA1} \\
R_{12}&=\log_2 \left(1+\frac{P_{12}g_{12}}{1+P_{11}g_{11}} \right). \label{eqn:MA2}
\end{align}
From (\ref{eqn:MA1}) and (\ref{eqn:MA2}), the transmit powers of $S_1$ and $S_2$ for a given channel
gain pair and transmit rate pair are derived in \cite{Chen2012two-way6} and are omitted here for brevity.

In addition, for Modes 2 to 4, we have
\begin{align}
   \bar{R}_i = E\{R_i(g_i)\}=E\{ \log_2(1+P_i(g_i)g_i) \}, \label{eq:rate}
\end{align}
\begin{align}
   \bar{P}_i = E\{P_i(g_i)\}= E\left\{ \frac{2^{R_i(g_i)}-1}{g_i} \right\}, \label{eq:power}
\end{align}
where $\bar{P}_i$ and $\bar{R}_i$ are averaged over the channel gain distribution.


As in the PNC based strategy, a fraction $f_i$ ($i = 1,\cdots,4$) of time is allocated to Mode $i$.
Therefore $f_i \bar{P}_i$ is proportional to the average energy used for transmission in Mode $i$.
The optimal DNC-TS strategy is found by solving problem {\bf P2}:
\begin{eqnarray}
   \min_{f_i,R_i(g_i)} && \sum_{i=1}^{4}f_i \bar{P}_i  \label{opt}
\end{eqnarray}
subject to
\begin{align}
f_{1}\bar{R}_{1i} &\ge  \lambda_i  \quad\quad i=1,2   \label{lopt_1}\\
f_2\bar{R}_2 + f_4\bar{R}_4 &\ge  \lambda_1    \label{lopt_2}\\
f_2\bar{R}_2 + f_3\bar{R}_3 &\ge \lambda_2    \label{lopt_3}\\
\sum_{i=1}^4 f_i &\leq 1   \label{lopt_4}
\end{align}

In \cite{Chen2012two-way}, it was observed that Mode 3 and Mode 4 are never both active due to network coding
gain. Hence assuming that $\lambda_1  \le \lambda_2$, Mode 4 can be dropped and only Mode 1 to Mode 3 will be used.
In addition, it was observed in \cite{Chen2012two-way} that
the rate constraints in (\ref{lopt_2}) and (\ref{lopt_3}) are met with equality. Hence we have
$\lambda_1 = f_2^* \bar{R}^*_2$ and $\lambda_2-\lambda_1 = f_3^* \bar{R}^*_3$.
Since {\bf P2} is a convex optimization problem, its optimal solution can be derived from the KKT conditions.
The details are presented in \cite{Chen2012two-way6} and are omitted here for brevity.
For Modes 2 to 3, we have a water-filling structure for the optimal power allocations:
\begin{equation}
 P_i^*(g_i) = \left[\beta^*_i \log_2 e -\frac{1}{g_i}\right]^+  i = 2,3. \label{eqn:kkt3}
\end{equation}

For Mode 1, resource allocation is however a little bit complicated and the result is summarized in the following lemma.
\begin{lemma}\label{uplink}
If $g_{11}>g_{12}$ and $\lambda_1 \le \lambda_2$, the power allocation for the multi-access
uplink transmission mode can be described as follows.
\begin{enumerate}
\item if $\beta_{12}^* \leq \beta_{11}^*$, then
\begin{align}
\left\{
\begin{array}{ll}
P_{11}^*=[\beta_{11}^*\log_2e-\frac{1}{g_{11}}]^+ \\
P_{12}^*=0 \\
\end{array}
\right.
\end{align}
\item if $\beta_{12}^* > \beta_{11}^*$, then
\begin{itemize}
\item{} if $\beta_{11}^*g_{11} \leq \beta_{12}^*g_{12}$, then
\begin{align}
\left\{
\begin{array}{ll}
P_{11}^*=0\\
P_{12}^*=[\beta_{12}^*\log_2e-\frac{1}{g_{12}}]^+\\
\end{array}
\right.
\end{align}
\item{} if $\beta_{11}^*g_{11}>\beta_{12}^*g_{12}$ and $\beta_{12}^*-\beta_{11}^* \leq \frac{g_{11}-g_{12}}{g_{11}g_{12}}$, then
\begin{align}
\left\{
\begin{array}{ll}
P_{11}^*=[\beta_{11}^*\log_2e-\frac{1}{g_{11}}]^+\\
P_{12}^*=0\\
\end{array}
\right.
\end{align}
\item{} if $\beta_{11}^*g_{11}>\beta_{12}^*g_{12}$ and $\beta_{12}^*-\beta_{11}^* > \frac{g_{11}-g_{12}}{g_{11}g_{12}}$, then
\begin{align}
\left\{
\begin{array}{ll}
P_{11}^*=[\frac{(\beta_{11}^*g_{11}-\beta_{12}^*g_{12})\log_2e}{g_{11}-g_{12}}]^+\\
P_{12}^*=[\frac{(\beta_{12}^*-\beta_{11}^*)g_{11}\log_2e}{g_{11}-g_{12}}-\frac{1}{g_{12}}]^+\\
\end{array}
\right.
\end{align}
\end{itemize}
\end{enumerate}
\end{lemma}
For the case that $g_{11}<g_{12}$ power allocation for Mode 1 can be derived in a similar manner.

\subsection{Digital Network Coding and Superposition in the Downlink (DNC-Sup)}\label{superposition}
The downlink channel from $R$ to $S_1$ and $S_2$ can be considered as a degraded broadcast channel,
except that each receiver has full knowledge of the message being transmitted to the other.
In Section II.C, the possibility of superposition coding at R and successive interference cancelation (SIC) decoding at $S_1$ and $S_2$ was
not considered. In this section, we will allow for the superposition of the network-coded message
with the remainder of the longer message in a new Mode 5, described as follows
when $\lambda_1 \le \lambda_2$, and when the message from $S_2$ is always longer than the one from $S_1$.
\begin{itemize}
\item{\em Mode 5:} $R$ broadcasts to $S_1$ and $S_2$ at the average rate pair
($\bar{R}_{51}, \bar{R}_{52}$) on the downlink. Each node decodes
the network coded message first and $S_1$ decodes the
remaining bits of the larger message after subtracting the network coded message.
$\bar{R}_{52}$ is the rate of the network coded message to both users and
$\bar{R}_{51}$ that
of the remaining bits of the longer message to $S_1$.
\end{itemize}

We now define $g_{51} = g_{r1}$, and $g_{52} = g_{r2}$, to be consistent
with the notation of previous sections.
If $g_{51}>g_{52}$, the required energy pair ($P_{51}$,$P_{52}$), for a given rate
pair ($R_{51}$,$R_{52}$), is given by,
\begin{align}
\left\{
\begin{array}{ll}
P_{51}(R_{51},R_{52})= \frac{2^{R_{51}}-1}{g_{51}} \\  
P_{52}(R_{51},R_{52})= \frac{2^{R_{52}}-1}{g_{52}}\left( 1+ \frac{g_{52}(2^{R_{51}}-1)}{g_{51}}\right) \\
\end{array} \label{eq:bc2}
\right.
\end{align}
Similarly, if $g_{51}<g_{52}$, the required energy pair ($P_{51}$,$P_{52}$), for a given rate pair ($R_{51}$,$R_{52}$) is given by,
\begin{align}
\left\{
\begin{array}{ll}
P_{51}(R_{51},R_{52})= \frac{2^{R_{51}}-1}{g_{51}} \\  
P_{52}(R_{51},R_{52})= \frac{2^{R_{51}}(2^{R_{52}}-1)}{g_{51}} \\
\end{array} \label{eq:bc3}
\right.
\end{align}

For a given gain pair $(g_{51},g_{52}) $, the total energy usage in
Mode 5 then is defined as $P_5=P_{51}+P_{52}$.

Note that in Section \ref{time sharing} Mode 5 was not allowed.
Introducing Mode 5 turns problem {\bf P2} into
\begin{eqnarray}
   \min_{f_i,R_i(g_i)} && \sum_{i=1}^{5}f_i \bar{P}_i  \label{P3_opt}
\end{eqnarray}
subject to
\begin{align}
f_{1}\bar{R}_{1i} &\ge \lambda_i   \quad\quad i=1,2   \label{P3_lopt_1}\\
f_2\bar{R}_2 + f_5\bar{R}_{52} &\ge \lambda_1   \label{P3_lopt_2}\\
f_3\bar{R}_3 + f_5\bar{R}_{52} &\le \lambda_2-\lambda_1    \label{P3_lopt_3}\\
\sum_{i=1}^5 f_i &\leq 1   \label{P3_lopt_4}
\end{align}

This new problem is equivalent to a simpler optimization problem with only Mode 1 and Mode 5 used. The conclusion is summarized below and the proof is given thereafter.

\begin{lemma} \label{lemma:equivalence}
The solution to (\ref{P3_opt}) must necessarily have only Mode 1 and Mode 5 active, i.e.,
\begin{align}
f_2^*=f_3^*=f_4^*=0,
\end{align}
where $f_i^*$ is the optimal value of $f_i$.
\end{lemma}
\begin{IEEEproof}
Note that from \cite{Chen2012two-way}, at most one of Mode 3 and Mode 4 will be active and this also applies to {\bf P2}. With the assumption that $\lambda_1<\lambda_2$, we conclude that $f_4^*=0$. It comes from the rate gain of network coding and hence we should employ network coding to the fullest extent in our strategy.

Below we prove $f_2^*=f_3^*=0$ by contradiction. Suppose in the optimal solution, we have both $f_2^*$ and $f_3^*$ positive. For a given gain pair in the downlink, ($g_{51}$,$g_{52}$), the total energy consumed in Mode 2 and Mode 3 is given by,
\begin{align}
E_1(g_{51},g_{52})=f_2^*\frac{2^{R_2^*}-1}{\min(g_{51},g_{52})}+f_3^*\frac{2^{R_3^*}-1}
{g_{51}}. \label{E1}
\end{align}

Now consider replacing Modes 2 and 3 with Mode 5. Since the bit rates in Modes 2 and 3 were $f_2^* R_2^*/(f_2^* + f_3^*)$ and $f_3^*R_3^*/(f_2^* + f_3^*)$ respectively, when Modes 2 and 3 are replaced by Mode 5, we must have broadcast-channel rates of
\begin{align}
R_{51}^{'}(g_{51},g_{52})=\frac{f_3^*R_3^*}{f_2^*+f_3^*},\\
R_{52}^{'}(g_{51},g_{52})=\frac{f_2^*R_2^*}{f_2^*+f_3^*}.
\end{align}
The corresponding total consumed energy for this gain pair per unit time is given by
\begin{align}
E_5^{'}=\left\{
\begin{array}{ll}
(f_2^*+f_3^* )\left(\frac{2^{R_{52}^{'}+R_{51}^{'}}-2^{R_{52}^{'}}}{g_{51}}+\frac{2^{R_{52}^{'}}-1}{g_{52}}\right) \hspace{0.05cm} \mbox{if $g_{51}\ge g_{52}$,} \\
(f_2^*+f_3^*)\frac{2^{R_{51}^{'}+R_{52}^{'}}-1}{g_{51}}  \quad \quad \mbox{Otherwise.}\\
\end{array}
\right.
\end{align}

We can now compare the energy consumption on the downlink for Mode 5 and time sharing of Mode2 and Mode 3. When $g_{51}
> g_{52}$, we have the comparison in (54)-(58), which shows that Mode 5 uses less average energy than
time sharing of Mode 2 and Mode 3.
\begin{figure*}[t]
\small
\begin{align}
E_5^{'}-E_1&=\left(f_2^*+f_3^* \right)\left(\frac{2^{R_{52}^{'}+R_{51}^{'}}}{g_{51}}-\frac{1}{g_{52}}-2^{R_{52}^{'}}\left(\frac{1}{g_{51}}-\frac{1}{g_{52}}\right) \right)-\left( f_2^*\frac{2^{R_2^*}-1}{g_{52}}+f_3^*\frac{2^{R_3^*}-1}
{g_{51}} \right) \label{eqn:1}\\
&=\left(f_2^*+f_3^* \right)\left(\frac{2^{\frac{f_2^*R_{2}^{*}+f_3^*R_{3}^{*}}{f_2^*+f_3^*}}}{g_{51}}-\frac{1}{g_{52}}-2^{R_{52}^{'}}\left(\frac{1}{g_{51}}-\frac{1}{g_{52}}\right) \right)-\left( f_2^*\frac{2^{R_2^*}-1}{g_{52}}+f_3^*\frac{2^{R_3^*}-1}
{g_{51}} \right) \label{eqn:11}\\
&<f_2^*\frac{2^{R_2^*}}{g_{51}}+f_3^*\frac{2^{R_3^*}}
{g_{51}}-\left( f_2^*\frac{2^{R_2^*}}{g_{52}}+f_3^*\frac{2^{R_3^*}}
{g_{51}} \right)-(f_2^*+f_3^*)2^{R_{52}^{'}}\left(\frac{1}{g_{51}}-\frac{1}{g_{52}}\right)
+f_3^*\left(\frac{1}{g_{51}}-\frac{1}{g_{52}}\right)\label{eqn:2} \\
&=\left(f_2^*2^{R_2^*}-(f_2^*+f_3^*)2^{R_{52}^{'}}\right)\left(\frac{1}{g_{51}}-\frac{1}{g_{52}}\right)+f_3^*\left(\frac{1}{g_{51}}-\frac{1}{g_{52}}\right) \label{eqn:3} \\
&=\left(f_2^*(2^{R_2^*}-1)-(f_2^*+f_3^*)(2^{R_{52}^{'}}-1)\right)\left(\frac{1}{g_{51}}-\frac{1}{g_{52}}\right)<0. \label{eqn:4}
\end{align}
\end{figure*}
 Note that (\ref{eqn:2}) follows from convexity and (\ref{eqn:4})
 follows from the fact
 that $f(t)=t(2^{\frac{a}{t}}-1)$ ($a>0,t>0$) is a strictly decreasing
 function of $t$ and $g_{51}>g_{52}$. Hence Mode 2 and Mode 3 should be replaced by Mode 5 in this case
to minimize energy usage.

On the other hand, if $g_{51}<g_{52}$, we have
\begin{align}
&E_5^{'}-E_1\\
=&\left(f_2^*+f_3^* \right)\frac{2^{R_{51}^{'}+R_{52}^{'}}-1}{g_{51}}
-f_2^*\frac{2^{R_2^*}-1}{g_{51}}+f_3^*\frac{2^{R_3^*}-1}
{g_{51}} \label{eqn:21}\\
=&\frac{\left(f_2^*+f_3^* \right)2^{\frac{f_2^*R_{2}^{*}+f_3^*R_{3}^{*}}{f_2^*+f_3^*}}-f_2^*2^{R_2^*}-f_3^*2^{R_3^*}}{g_{51}}<0 \label{eqn:22}
\end{align}
where (\ref{eqn:22}) follows from convexity.

Hence we have proved that for any channel gain pair we can
use Mode 5 to replace Mode 2 and Mode 3 to reduce energy consumption
while delivering the same rates. Averaging over all possible channel
gain pairs, it is directly deduced that the average energy
consumed by employing Mode 5 is less than that with Mode 2 and Mode 3.
The assumed optimal solution to {\bf P2} thus could not be optimal.
\end{IEEEproof}

\textit{Remarks}: Intuitively,
with SIC decoding allowed at each receiver, DNC-Sup is able to achieve all points in the capacity region
of the broadcast channel while DNC-TS can not. Therefore DNC-Sup outperforms DNC-TS.

Henceforth, an equivalent optimization problem which only consists of Mode 1 and Mode 5, {\bf P2}',
aiming to minimize total average
energy usage,
is formulated as follows.
\begin{align}
\min_{f_i,R_i(g_i)} f_1\bar{P}_1+f_5\bar{P}_5
\end{align}
subject to
\begin{align}
f_1\bar{R}_{1i} &\geq \lambda_i \quad i=1,2\\
f_5\bar{R}_{51} &\geq \lambda_2-\lambda_1   \label{eq:excess}\\
f_5\bar{R}_{52} &\geq \lambda_1    \label{eq:common}\\
f_1 + f_5 &\leq 1
\end{align}

The optimal solution of {\bf P2'} is straightforward to obtain and was presented in [8].
Note that the optimal power allocation in both modes are identical to those of Mode 1 and Mode 3
in the PNC-based strategy.

It is noted that {\bf P2} and {\bf P2}' differ only in the downlink
transmission.
The former time shares the digital network-coded bits and the remaining bits of the larger message
in the downlink and the latter superimposes the network-coded bits
and the remaining bits of the larger message in the downlink.

\subsection{Codeword Superposition in the Downlink (CW-Sup)}\label{other}
In \cite{Rankov} and \cite{Boche}, the authors discussed a codeword superposition of the two original messages from $S_1$ and $S_2$ at $R$ for transmission in the downlink. As each source has the knowledge of the message transmitted by itself, it can subtract this message and decode the intended message without interference. Hence the broadcast channel in the downlink is equivalent to two interference-free AWGN channels.

Note that in this codeword superposition strategy, the relay node simply superimposes the two messages from the two sources and no network coding technique is applied. The transmit power of the relay node then is the sum of the transmit power for transmission of each of the two messages to the intended source. For convenience,
we refer to this transmission strategy in the downlink as Mode 6
and we shall analyze Mode 6 in our setup and minimize the total energy usage.
Note that in \cite{Rankov}\cite{Boche}, the authors assumed that each node is subject
to individual power constraint and focused on the throughput region of this network.
In our setup, however, we are more interested in total energy usage of this system instead
of deriving the throughput region.

\begin{itemize}
\item{\em Mode 6:} On average, $R$ broadcasts to $S_1$ and $S_2$ at rate pair
($\bar{R}_{61}, \bar{R}_{62}$) in the downlink. Each node subtracts the message originating from itself
and then decodes the intended message from the other source.
\end{itemize}

For a given rate pair (${R}_{61}, {R}_{62}$) and gain pair ($g_{61}$,$g_{62}$), the energy required is given by
 \begin{align}
 P_{6i}(g_{6i})&=\frac{2^{R_{6i}}-1}{g_{6i}} \quad i=1,2, \label{Boche1}
 \end{align}
where each source enjoys an interference-free Gaussian channel with SIC within one slot. Hence
the capacity on either side in the downlink is given by the Shannon
formula $C=\log_2(1+P_{6i}g_{6i})$ ($i=1,2$) and the total power required is $P_6=P_{61}+P_{62}$.
In this section, we are interested in deriving the minimal energy usage with Mode 1 and Mode 6. We are further interested in comparing energy usage of the designed transmission strategy in the previous section with the one consisting of Mode 1 and Mode 6.

The optimization problem to minimize energy usage with Mode 1 and Mode 6 can be formulated as {\bf P3},
\begin{align}
\min_{f_i,R_i(g_i)} f_1\bar{P}_1+f_6\bar{P}_6
\end{align}
subject to
\begin{align}
f_1\bar{R}_{1i} &\geq \lambda_i \quad i=1,2  \label{eq:up6}\\
f_6\bar{R}_{6i} &\geq \lambda_{3-i}  \quad i=1,2 \label{eq:down6}\\
f_1 + f_6 &\leq 1
\end{align}

Since {\bf P5} is a convex optimization problem, we can derive KKT conditions and obtain the optimal solution. Hence, the optimal solution to {\bf P5} has
\begin{align}
 &P_{6i}^*(g_{6i}) = \left[ \beta^*_{6i} \log_2 e -\frac{1}{g_{6i}} \right]^+,   \quad  i=1,2   \label{eqn:6kkt3}
\end{align}
where $\beta^*_{6j}$ ($j=1,2$) are Lagrangian multipliers.
It is observed in (\ref{eqn:6kkt3}) that the optimal power allocation again has a water-filling structure. Note that the optimal power allocation for the uplink
is identical to Lemma \ref{uplink} and is hence omitted.

Comparing DNC-Sup and CW-Sup, we can prove the following lemma.
\begin{lemma} \label{codeword}
For ergodic fading channels, DNC-Sup performs no worse than CW-Sup in terms of energy usage,
which indicates the superiority of our proposed transmission strategy in the downlink.
\end{lemma}
\begin{IEEEproof}
Suppose in the optimal solution to {\bf P5}, the optimal rate pair for a given gain pair ($g_{61}$,$g_{62}$) in the downlink is $R_{61}^*$ and $R_{62}^*$ and the energy required is given in (\ref{Boche1}).

We construct another solution which employs Mode 5 with the optimal assigned time fraction $f_6^*$.

Comparing the link gains $g_{61}$ and $g_{62}$ and the associated transmit rates $R_{61}^*$ and $R_{62}^*$, there are four cases to be investigated.

\begin{itemize}
\item{Case i)}: $g_{61}>g_{62} $ and $R_{61}^*>R_{62}^*$.
\end{itemize}

In Case i), by employing Mode 5, we have
\begin{align}
R_{51}^{'}(g_{51},g_{52})&= R_{61}^*-R_{62}^*, \\
R_{52}^{'}(g_{51},g_{52})&= R_{62}^*.
\end{align}
The energy required is given by,
\begin{align}
E_5^{'}&=\frac{2^{R_{52}^{'}+R_{51}^{'}}}{g_{51}}-\frac{1}{g_{52}}-2^{R_{52}^{'}}\left(\frac{1}{g_{51}}-\frac{1}{g_{52}}\right)\\
&=\frac{2^{R_{61}^*}}{g_{61}}-\frac{1}{g_{62}}
-2^{R_{62}^*}\left(\frac{1}{g_{61}}-\frac{1}{g_{62}}\right).
\end{align}

Hence we have
\begin{align}
E_5^{'}-E_6^*&=\frac{1}{g_{61}}-2^{R_{62}^*}\frac{1}{g_{61}} \le 0
\end{align}
since $2^{R_{62}^*} \ge 1$.

\begin{itemize}
\item{Case ii)}: $g_{61}>g_{62}$ and $R_{61}^*<R_{62}^*$.
\end{itemize}

In this case, we can simply transmit a network coded message at rate $R_{61}^*$, in other words, we have
\begin{align}
R_{51}^{'}(g_{51},g_{52})&= 0, \\
R_{52}^{'}(g_{51},g_{52})&= R_{61}^*.
\end{align}
Both sources can obtain $R_{61}^*$ bits of message, hence $S_2$ can obtain more information than in Mode 6. The energy required in Mode 5, is given by
\begin{align}
E_5^{'}=\frac{2^{R_{61}^*}-1}{g_{62}}<
\frac{2^{R_{62}^*}-1}{g_{62}}+\frac{2^{R_{61}^*}-1}{g_{61}}=E_6^*.
\end{align}

\begin{itemize}
\item{Case iii)}. $g_{61}<g_{62}$ and $R_{61}^*>R_{62}^*$.
\end{itemize}

In Case iii), by employing Mode 5, we have
\begin{align}
R_{51}^{'}(g_{51},g_{52})&= R_{61}^*-R_{62}^*, \\
R_{52}^{'}(g_{51},g_{52})&= R_{62}^*.
\end{align}
The energy required is given by,
\begin{align}
E_5^{'}=\frac{2^{R_{51}^{'}+R_{52}^{'}}-1}{g_{61}}=\frac{2^{R_{61}^*}-1}{g_{61}}.
\end{align}

Hence we have
\begin{align}
E_5^{'}=\frac{2^{R_{61}^*}-1}{g_{61}}<
\frac{2^{R_{62}^*}-1}{g_{62}}+\frac{2^{R_{61}^*}-1}{g_{61}}=E_6^*.
\end{align}

\begin{itemize}
\item{Case iv)}. $g_{61}<g_{62}$ and $R_{61}^*<R_{62}^*$.
\end{itemize}

In this case, we can also simply transmit a network coded message at rate $R_{62}^*$, in other words, we have
\begin{align}
R_{51}^{'}(g_{51},g_{52})&= 0, \\
R_{52}^{'}(g_{51},g_{52})&= R_{62}^*.
\end{align}
Both sources can obtain $R_{62}^*$ bits of message, hence $S_1$ can obtain more information than in Mode 6. The energy required in Mode 5, is given by
\begin{align}
E_5^{'}=\frac{2^{R_{62}^*}-1}{g_{61}}<
\frac{2^{R_{62}^*}-1}{g_{62}}+\frac{2^{R_{61}^*}-1}{g_{61}}=E_6^*.
\end{align}

Hence we have verified that for all cases Mode 5 performs better than Mode 6 in terms of energy usage. 
\end{IEEEproof}

\subsection{The Optimal Scheme}
We have so far analyzed four different strategies for TWRNs.
Here we design an algorithm to obtain the optimal solution among
these strategies for TWRNs with arbitrary rate pair requirements. Note that we have shown that DNC-Sup outperforms DNC-TS and CW-Sup. Hence only PNC-Sup and DNC-Sup are considered in the optimal solution in terms of energy usage, which is given as follows.
\begin{enumerate}
\item For a given rate pair requirement, we solve {\bf P1} for PNC-Sup and {\bf P2'} for DNC-Sup to obtain their total energy usage.
\item We compare the total energy usage of PNC-Sup and DNC-Sup. If PNC-Sup uses less energy, it is selected as the optimal transmit strategy. Otherwise, DNC-Sup is selected.
\end{enumerate}
In this way, we always select the optimal strategy and the minimum total average energy usage is then achieved for arbitrary rate pair requirements. For reference, we call it {\bf Popt}.

\section{Numerical Results}
We now present numerical results to verify our findings.
Noise at each node is assumed to be Gaussian with zero mean and unit variance and
all links are assumed to be Rayleigh fading channels.
The associated instantaneous channel state information (CSIT), as well as channel statistics are assumed
to be known to the corresponding transmit nodes.
Moreover, coding and decoding algorithms are not employed directly.
Rather, we use the theoretical achievable rate over each link as the transmit rate. It is also noted that the unit of
rate requirement on either side is frames per slot. In addition,
the constant overhead energy usage (for circuit operation) for a TWRN is assumed to be zero,
as it does not affect
the transmit energy usage for the different strategies.

For comparison with {\bf P1}, we also present the achievable minimal average energy used for a TWRN
by PNC with zero padding (PNC-ZP). Specifically, when the messages from the two sources are not of equal length, we zero pad the shorter
message to make it equal in length to the longer message. The associated optimization problem is referred
to as {\bf P0}:
\begin{align}
\min_{g_{ij}} \quad \sum_{i=1}^2 f_i^{\mathrm{PNC}}\bar{P}_i^{\mathrm{PNC}} \label{eqn:pnczpobj}
\end{align}
subject to the following constraints,
\begin{align}
&f_1^{\mathrm{PNC}}\bar{R}_1^{\mathrm{PNC}} \ge \max(\lambda_1,\lambda_2) \label{eqn:pnczpcon1}\\
&f_2^{\mathrm{PNC}}\bar{R}_2^{\mathrm{PNC}} \ge \max(\lambda_1,\lambda_2) \label{eqn:pnczpcon2}\\
&P_{11}^{\mathrm{PNC}}g_{1r}=P_{12}^{\mathrm{PNC}}g_{2r} \label{eqn:pnczpcon3}\\
&f_1 + f_2  \leq 1   \label{eqn:pnczpcon4}
\end{align}
where (\ref{eqn:pnczpcon1}) and (\ref{eqn:pnczpcon2}) follows
from the fact that the two transmitted messages have the same length after zero-padding.
$f_1$ is the time fraction assigned for the transmission on the uplink and $f_2$ is for transmission over the downlink.
This is a standard convex optimization problem and the solution to {\bf P0} is given by,
\begin{align}
P_{1i}^{\mathrm{PNC}}(g_{1r},g_{2r})&=\left[\beta_1^* \log_2 e \frac{g_{3-i,r}}{g_{1r}+g_{2r}}-\frac{1}{2g_{ir}}\right]^+, \label{eqn:pnzpcopt2}\\
P_2^{\mathrm{PNC}}(g_{r1},g_{r2})&=\left[\beta_2^*\log_2 e - \frac{1}{g_{2r}}\right] ,\label{eqn:pnzpopt3}
\end{align}
where $\beta_1$ and $\beta_2$ are Lagrangian multipliers for (\ref{eqn:pnczpcon1}) and (\ref{eqn:pnczpcon2}) respectively.

For clarity, Table \ref{Tab1} lists the different optimization problems with the corresponding
transmit strategies.
\begin{table}[h!] 
\caption{list of Different Optimization problems with their adopted strategies}
\centering
\begin{tabular}{|c|c|m{3.7cm}|}
ine
Problem Index &  Strategy Employed  &  Detailed Description  \\
ine
{\bf P0} & PNC-ZP & Physical layer network coding with zero padding for the smaller message in the downlink \\
ine
{\bf P1} & PNC-Sup & Physical layer network coding with superposition in the downlink \\
ine
{\bf P2} & DNC-TS & Time sharing of DNC message and the remaining bits of the larger message in the downlink \\
ine
{\bf P2}' & DNC-Sup & Superposition of DNC message and the remaining bits of the larger message in the downlink\\
ine\
{\bf P3} & CW-Sup & Superposition of two original codeword messages in the downlink\\
ine
\end{tabular}
\label{Tab1}
\end{table}

\begin{figure}[t]
   \centering
   \includegraphics[width = 12cm]{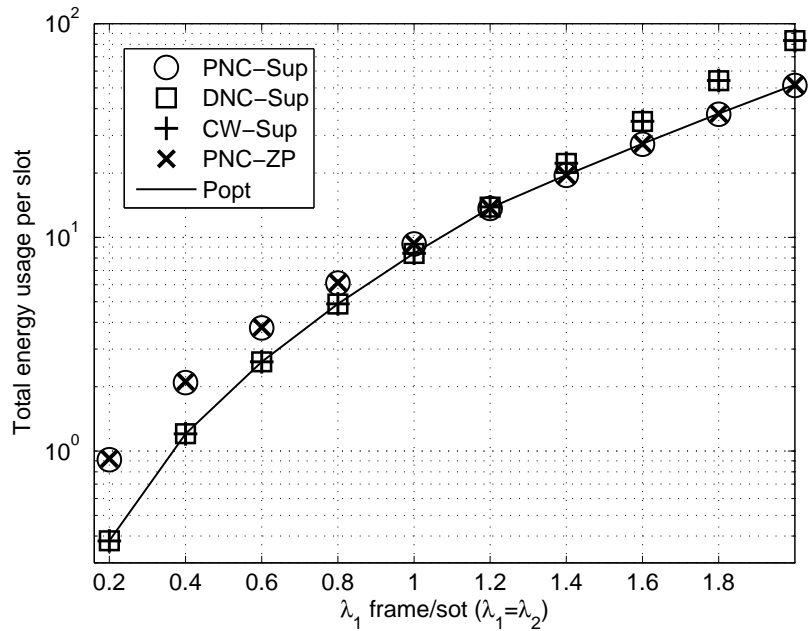}
   \caption{Comparison of total energy consumption for the optimal
   solutions to {\bf Popt}, {\bf P1}, {\bf P2}',
   {\bf P3} where we set $\bar{g}_{1r}=\bar{g}_{2r}=\bar{g}_{r1}=1$ and $\bar{g}_{r2}=2$.
   We also assume symmetric data rate requirements for both sources.} \label{fig:energy_usage_strategy1}
   \end{figure}

In Fig. \ref{fig:energy_usage_strategy1}, the minimal average transmit energy usage
per slot with symmetrical rate requirements ($\lambda_1=\lambda_2$) of different strategies are compared.
Under a symmetric traffic scenario, DNC-TS and DNC-Sup are identical to each other, hence only
the solution to DNC-Sup is plotted. It is also seen that PNC-ZP performs identical to PNC-Sup
as they are the same for symmetric traffic case. It is observed that PNC-Sup
outperforms DNC-Sup and CW-Sup at relatively high data
rate requirements, i.e., $\lambda_i>1.2$ frame/slot. However, with low data rate requirements,
DNC-Sup and CW-Sup perform better than PNC-Sup in terms of total energy usage. It is because, with low
data rate requirement, interference in joint decoding in the multi-access uplink plays a negligible
role in degrading performance. However, with high data rate requirements, interference from joint-decoding
in the multi-access uplink dominates the performance of the uplink transmission with
DNC-Sup and CW-Sup. On the other hand, PNC-Sup can perform better in
the high data rate requirement regime, because the relay node only decodes a function of
individual messages in the uplink instead of jointly decoding two messages. It is also interesting
to note that DNC-Sup and CW-Sup are quite close to each other with symmetric traffic,
which intuitively follows from DNC-Sup being bounded by the instantaneous minimum gain of
two downlink channels and CW-Sup suffering from more messages
being transmitted in the downlink. As designed, {\bf Popt} always selects the optimal strategy in terms of energy usage and performs best for all rate pair requirements.


We next discuss the performance of the different strategies for asymmetric traffic scenarios ($\lambda_1 \neq \lambda_2$).

\begin{figure}[t]
   \centering
   \includegraphics[width = 12cm]{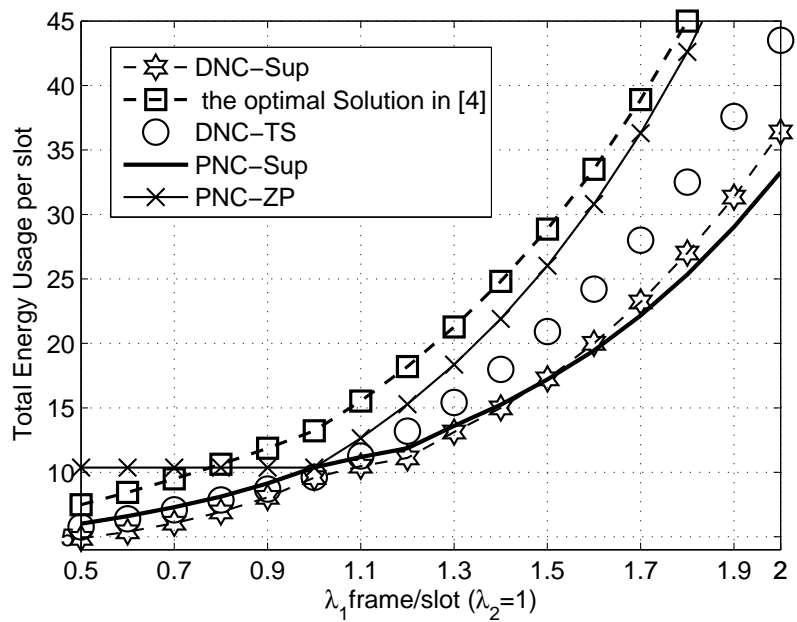}
   \caption{Comparison of optimal energy consumption for the optimal solution
   in \cite{Chen2012two-way}, {\bf P0}, {\bf P1}, {\bf P2} and {\bf P2}' for asymmetric traffic scenarios, where we set $\bar{g}_{1r}=\bar{g}_{2r}=\bar{g}_{r1}=\bar{g}_{r2}=1$.} \label{fig:energy_usage_bc}
   \end{figure}

In Fig. \ref{fig:energy_usage_bc}, we compare
the optimal time-sharing solution in \cite{Chen2012two-way},
PNC-ZP, PNC-Sup, DNC-TS and DNC-Sup.
It is observed that with multi-access transmission, DNC-TS performs better than the solution in
\cite{Chen2012two-way}
in terms of energy usage, as \cite{Chen2012two-way} only considers orthogonal, time-sharing
transmissions, which was investigated by Yeung in terms of achievable throughput region in
\cite{fong2011practical}.
Hence our strategy also outperforms the strategy in \cite{fong2011practical} in terms of energy usage.

It is observed that with superposition coding in the downlink by DNC-Sup
for asymmetric traffic,
we can perform even better in terms of energy, which validates Lemma \ref{lemma:equivalence}.
It should also be noted that with $\lambda_1$ and $\lambda_2$ approaching each other,
the energy benefit from the superposition coding on the downlink gradually decreases
as fewer remaining bits can be superimposed on the network coded message.
It is interesting to note that the minimal energy usage by PNC-ZP
is a constant when $\lambda_1<\lambda_2$ and that it performs worse than
PNC-Sup, which is due to the fact that with zero padding
the virtual traffic is determined by $\max(\lambda_1,\lambda_2)$ and will
incur unnecessary energy usage.

\begin{figure}[t]
   \centering
   \includegraphics[width = 12cm]{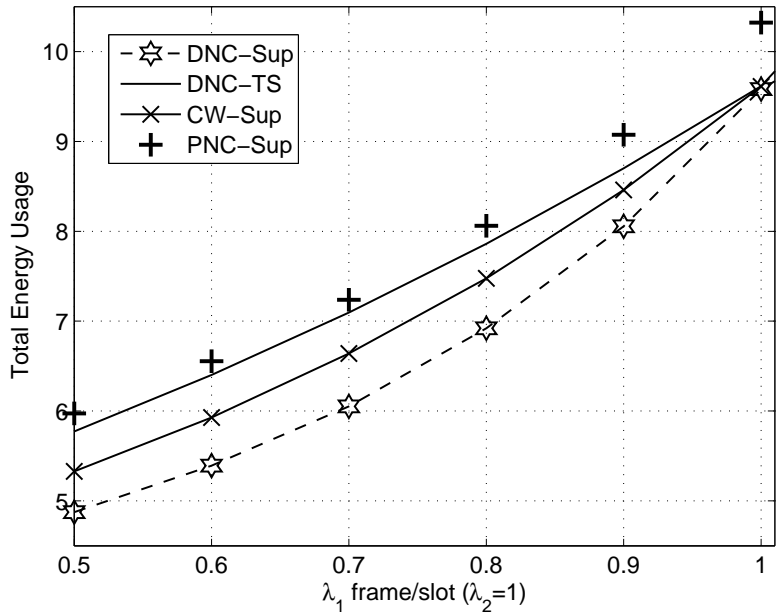}
   \caption{Comparison of optimal energy consumption for {\bf P1}, {\bf P2}, {\bf P2}' as well as {\bf P3}
   for asymmetric traffic scenarios. All links are assumed to have unity gain on average.} \label{fig:energy_comparison_boche}
   \end{figure}

In Fig. \ref{fig:energy_comparison_boche}, we compare PNC-Sup, DNC-TS,
DNC-Sup and CW-Sup.
It is observed that with small rate pair requirements, PNC-Sup performs worse than the other strategies.
It is also observed that CW-Sup is worse than DNC-Sup in terms of energy usage, which verifies
our theoretical observation in Lemma \ref{codeword}. However, it is noted that DNC-TS,
which employs time sharing in the downlink, consumes more energy resources than CW-Sup.
This intuitively follows from two facts. The first is that the network coded message enjoys a
channel whose average link gain is less than that of either link in the downlink,
while in CW-Sup, messages are sent over interference-free individual channels whose
average link gains are unity. The second is that in CW-Sup,
both messages can
use more time resources for transmission in the downlink, whereas in DNC-TS,
the network-coded message and the excess bits of the larger message
compete for time-resource allocation.

\begin{figure}[t]
   \centering
   \includegraphics[width = 12cm]{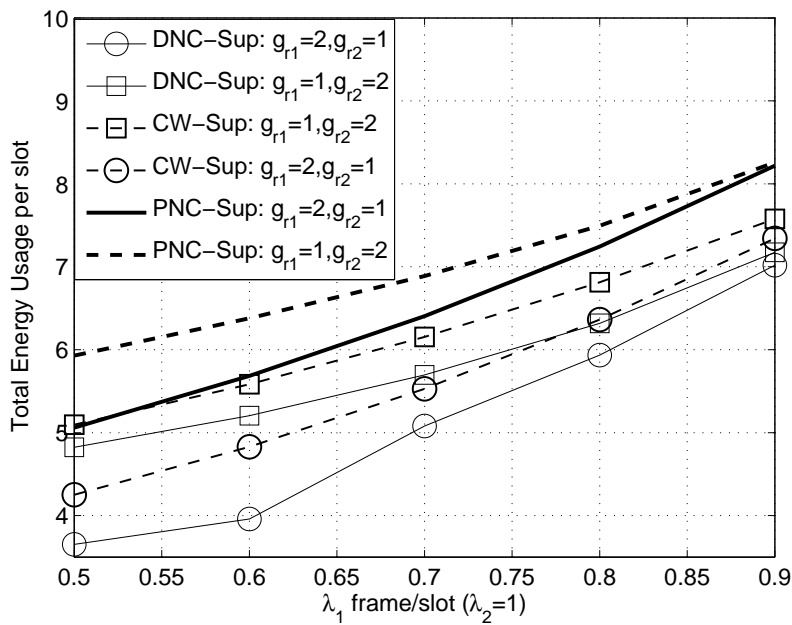}
   \caption{Comparison of optimal transmit energy consumption for different strategies with different average channel gains
   in the downlink for asymmetric traffic scenarios. Two cases are taken into consideration.
   One is that $\bar{g}_{1r}=\bar{g}_{2r}=\bar{g}_{r2}=1$ and $\bar{g}_{r1}=2$. The other is that
   $\bar{g}_{1r}=\bar{g}_{2r}=\bar{g}_{r1}=1$ and $\bar{g}_{r2}=2$.} \label{fig:unequal_gain}
   \end{figure}

In Fig. \ref{fig:unequal_gain}, we compare
DNC-Sup and CW-Sup for different average channel gain pairs.
It can be observed that the total average energy used per slot in the case of
$\bar{g}_{r1}=1,\bar{g}_{r2}=2$ is less than that in the case
of $\bar{g}_{r1}=2,\bar{g}_{r2}=1$ for both DNC-Sup and CW-Sup strategies.
For DNC-Sup, this is because the energy consumption for the remaining bits of the larger message
is determined by $\bar{g}_{51}$, i.e., $\bar{g}_{r1}$ and the case with $\bar{g}_{r1}=2$ obviously
performs better than the case with $\bar{g}_{r1}=1$. For CW-Sup, it is because the
larger message is transferred over the link $R-S_2$ on the downlink.
Hence, the performance of CW-Sup improves with higher $\bar{g}_{2r}$.

\section{Conclusion}
In this work, the problem of minimizing energy usage in a TWRN over a fading channel was formulated and solved for various transmit strategies and comparisons were performed.
Three transmission strategies were considered: physical-layer network coding (PNC),
digital-network coding (DNC), and codeword superposition (CW-Sup).
In the downlink for DNC, a simple time sharing strategy of the digital network-coded message and the
remaining bits of the larger message (DNC-TS) was first considered and extended to
a superposition strategy which superimposed these two messages (DNC-Sup).
Between DNC and CW-Sup,
the superiority of the superposition of network coded message and the
excess bits of the larger message (DNC-Sup), was demonstrated theoretically
in terms of energy usage. More importantly, it was shown that, in terms of total energy usage,
the specific PNC scheme performs better than DNC in the regime of relatively high data rate requirements,
 and worse than DNC with relatively low data rate requirements. This provides some insights on when to select PNC or DNC
 for TWRNs.
 Finally, an optimal algorithm to always select the best strategy
 in terms of energy usage was presented.

\end{document}